\pdfoutput=1
\documentclass[10pt,conference]{IEEEtran}

\usepackage{booktabs}
\usepackage{multirow}
\usepackage{amssymb}
\usepackage{pifont}
\newcommand{\xmark}{\ding{55}}
\usepackage{array}
\newcolumntype{L}[1]{>{\raggedright\arraybackslash}p{#1}}
\usepackage{amsmath}
\usepackage{algorithm}
\usepackage[noend]{algpseudocode}
\algnewcommand{\ParFor}[1]{\State \textbf{in parallel for} #1 \textbf{do}}
\algnewcommand{\EndParFor}{\State \textbf{end parallel for}}
\algrenewcommand{\algorithmiccomment}[1]{\hfill\textcolor{gray}{\textit{// #1}}}
\algnewcommand{\PhaseBar}[1]{\Statex\hspace*{-\fboxsep}\colorbox{gray!22}{\makebox[\dimexpr\linewidth-2\fboxsep][l]{\textit{#1}}}}
\usepackage{graphicx}
\usepackage{url}
\usepackage{listings}
\usepackage{xcolor}
\usepackage{tikz}
\usepackage{tcolorbox}
\tcbuselibrary{listings,skins}
\usetikzlibrary{arrows.meta,positioning,calc,shadows}
\graphicspath{{figures/}}

\definecolor{listingbackground}{HTML}{F5F8FC}
\definecolor{listingrule}{HTML}{AEBED0}
\definecolor{listingkey}{HTML}{245B8A}
\definecolor{listingstate}{HTML}{694A9C}
\definecolor{listingsuccess}{HTML}{1D7064}
\definecolor{listingfailure}{HTML}{A33A3A}
\definecolor{listingarrow}{HTML}{52687D}
\definecolor{listingcomment}{HTML}{667788}
\definecolor{searchboxframe}{HTML}{74A39B}
\definecolor{searchboxtitle}{HTML}{275F58}
\definecolor{constructionboxframe}{HTML}{9585AF}
\definecolor{constructionboxtitle}{HTML}{594876}
\newif\ifshownotes
\shownotesfalse
\definecolor{authornote}{HTML}{B03A2E}

\definecolor{mpx}{HTML}{264aac}

\providecommand{\Description}[1]{}
\renewcommand{\paragraph}[1]{\par\emph{#1}\hspace{0.5em}\ignorespaces}

\newcommand{\code}[1]{\texttt{\small #1}}

\begin{document}

\title{BenchMeUp: Building Software Engineering Benchmarks from Natural-Language Specifications}

\author{\IEEEauthorblockN{Islem Bouzenia}
\IEEEauthorblockA{CISPA Helmholtz Center for Information Security\\
Germany\\
bouzenia.islem@pm.me}
\and
\IEEEauthorblockN{Michael Pradel}
\IEEEauthorblockA{CISPA Helmholtz Center for Information Security\\
Germany\\
michael@binaervarianz.de}}

\maketitle

\begin{abstract}
Software-engineering research relies on benchmarks that evaluate new techniques on a set of problem instances.
Unfortunately, manually constructing benchmarks is labor-intensive, as researchers must find suitable instances, prepare execution environments, establish expected outcomes, and validate correctness.

This paper presents BenchMeUp, an approach for automatically creating software-engineering benchmarks from natural-language specifications.
Given a specification defining the desired instances, data sources, and evaluation procedures, BenchMeUp generates search rules and a construction plan, invokes parallel LLM agents that identify candidate instances and construct executable instances, validates the results in fresh execution environments, and packages the instances behind a unified command-line interface.
We evaluate the approach on four tasks: program repair, issue reproduction, compiler autotuning, and fixing static-analysis warnings.
Targeting 50 instances per task (200 in total), BenchMeUp constructs 198 instances, of which 196 (99\%) are successfully packaged.
Search and construction take 36 minutes and \$0.79 per instance, on average.
Evaluating eight existing techniques on the generated benchmarks, we find that 189 of 196 packaged instances (96\%) support end-to-end evaluation, allowing users to study the effectiveness of the techniques.
Checking every constructed instance against its source, the specification, and its expected outcome, we find that 180 of 198 (91\%) meet all criteria, compared with 10 of 200 instances delivered by Codex (general-purpose coding agent).

Overall, we show that LLM-based agents can automate benchmark construction across diverse software-engineering tasks and produce executable instances that existing techniques can use, while validity beyond execution still needs to be checked.

\end{abstract}


\section{Introduction}
\label{sec:intro}

Benchmarks are central to software-engineering research~\cite{AI4SEBenchmarkDesign2026}, enabling researchers to evaluate techniques on common problem instances, compare results, reproduce previous findings, and measure progress. Some benchmarks have shaped entire research areas, such as Defects4J~\cite{just2014defects4j} for automated program repair~\cite{cacm2019-program-repair} and SWE-bench~\cite{jimenez2024swebench} for evaluating LLM-based issue-resolution agents.

Unfortunately, manually constructing benchmarks is labor-intensive.
Researchers must first define suitable benchmark instances and identify their data sources. They must then collect and prepare instance artifacts, such as source code and tests, set up dependencies, determine expected outcomes, and verify that each instance meets the benchmark requirements. For benchmarks involving executable software, this process often requires writing task-specific scripts, building and configuring projects, and manually inspecting the results.
This effort leads researchers to reuse benchmarks for long periods, raising concerns about overfitting and representativeness~\cite{AI4SEBenchmarkDesign2026}.
Exploring new tasks or programming languages also requires substantial effort to construct suitable benchmarks.

Several approaches automate parts of benchmark construction. BugSwarm~\cite{tomassi2019bugswarm} mines continuous-integration histories to collect reproducible failing and passing program versions. SetUpAgent~\cite{vergopoulos2025setupagent}, ExecutionAgent~\cite{bouzenia2025you}, and RepoLaunch~\cite{li2026repolaunch} automate building and testing repositories. SWE-smith~\cite{yang2025swesmith}, SWE-rebench~\cite{badertdinov2025swerebench}, SWE-Factory~\cite{guo2026swefactory}, SWE-Bench++~\cite{wang2025swebenchpp}, and SEC-bench~\cite{lee2025secbench} generate or collect issue-resolution and vulnerability tasks with executable environments. Each of these approaches targets one task or automates one construction step.
Beyond software engineering, AutoBencher~\cite{li2025autobencher} and BenchAgents~\cite{butt2024benchagents} construct benchmarks from natural-language descriptions, but focus on synthetic questions for evaluating language models.
To our knowledge, no existing approach creates real-world, executable benchmarks from natural-language specifications.

Recent advances in LLM-based agents~\cite{yang2024sweagent,bouzenia2025you,li2026repolaunch} raise the question of whether they can construct software-engineering benchmarks.
However, our evaluation shows that simply asking a general-purpose agent to do so introduces several challenges.
First, agents may interpret an ambiguous benchmark description inconsistently, selecting instances by different criteria and producing an incoherent benchmark.
For example, when collecting repair instances, one agent session includes style changes or feature implementations as ``bugs'', while another accepts only changes that correct behavioral faults revealed by tests.
Second, agents may construct instances that are difficult to reproduce.
For example, an agent setting up an open-source project relies on preinstalled dependencies, producing an instance that works in its environment but fails on another machine.
Finally, agents may expose different instances via different interfaces, even when instructed to use a common interface.
For example, an agent's \code{run\_tests.sh} executes all tests for one instance but only the bug-exposing test for another.
Hence, researchers cannot evaluate a technique reliably and uniformly across such instances.

This paper presents BenchMeUp, an approach for constructing software-engineering benchmarks from a researcher-written \emph{benchmark specification}. The specification describes the task, desired instances, their sources, and how to use them to evaluate a technique.
For example, a program-repair specification may request real-world bugs from open-source projects, each with a buggy program and a test exposing the bug. Researchers can specify project properties, such as language, development activity, and popularity, without knowing which repositories contain suitable instances.

BenchMeUp constructs a benchmark from the specification in four phases. First, an LLM generates explicit \emph{search rules} for finding and selecting candidates, followed by a shared \emph{construction plan} defining the construction steps and expected artifacts. Second, LLM-based \emph{search agents} autonomously find and record candidates satisfying the rules. Third, \emph{construction agents} follow the plan to build instances from the recorded candidates; BenchMeUp validates the instances by replaying the generated scripts in a fresh environment. Finally, BenchMeUp packages completed instances behind a unified command-line interface.
This design aims for consistent instance selection, reproducible outcomes in fresh environments, and a unified interface for evaluating techniques.

We evaluate our approach on four software-engineering tasks: program repair, issue reproduction, compiler autotuning, and fixing static-analysis warnings. Requesting 50 instances per task, we obtain 198 constructed instances, of which 196 are successfully packaged. Search and construction average 36 minutes and \$0.79 in LLM costs per instance.
We evaluate eight existing techniques across the four tasks using the constructed benchmarks. The techniques can set up and run 189 of the 196 packaged instances (96\%), enabling comparisons on common instances. Successful execution, however, does not establish validity. Checking every constructed instance against its source, the specification, and its expected outcome, we find that 180 of 198 (91\%) meet all criteria. Most remaining instances concern issue reproduction and fail the specification's filters or expected outcome. Manual inspection of 40 instances also reveals six with defective check scripts.
By comparison, only 10 of the 200 instances delivered for the same specifications by Codex~\cite{openai2025codex}, a general-purpose coding agent, meet all criteria.

In summary, this paper makes the following contributions:

\begin{itemize}

    \item \emph{Idea}. We propose creating arbitrary software-engineering benchmarks from natural-language specifications.
    
    \item \emph{Technique}. We present BenchMeUp, an automated, LLM-based, multi-agent approach that translates a natural-language specification into search rules and a construction plan. Agents follow these to construct benchmark instances, which are validated in fresh execution environments and packaged behind a common interface.
    
    \item \emph{Empirical evidence}. We evaluate the approach on four software-engineering tasks, compare BenchMeUp with a general-purpose agent, and apply eight existing techniques to the automatically constructed benchmarks.
    
    \item \emph{Open science}. \url{https://github.com/sola-st/BenchMeUp}
\end{itemize}

\section{Approach}
\label{sec:approach}
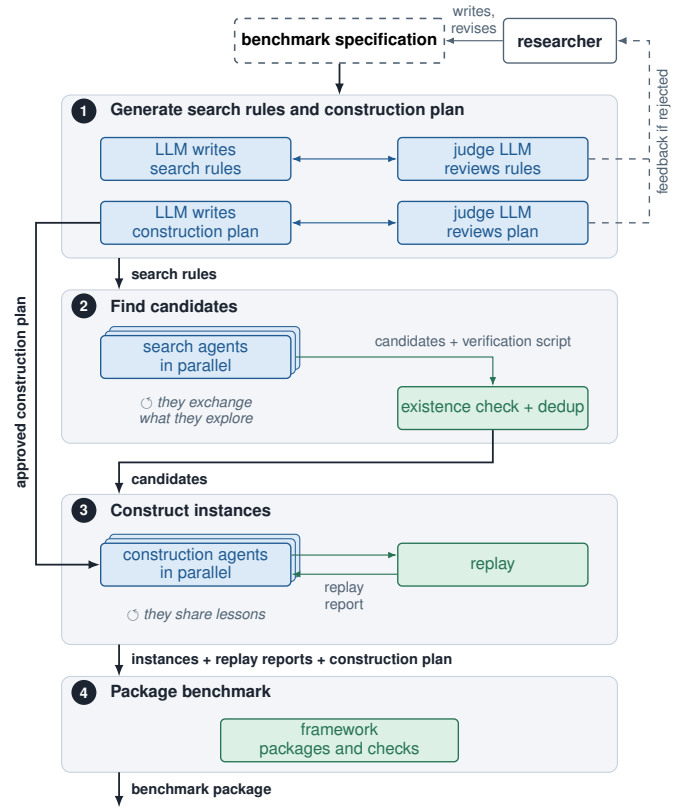
\begin{figure}[t]
\centering
\definecolor{ovblue}{HTML}{1F5A91}
\definecolor{ovbluebg}{HTML}{DCEAF7}
\definecolor{ovgreen}{HTML}{267355}
\definecolor{ovgreenbg}{HTML}{DCF0E5}
\definecolor{ovgray}{HTML}{506273}
\definecolor{ovpanel}{HTML}{F3F5F8}
\definecolor{ovpanelline}{HTML}{C9D3DE}
\definecolor{ovdark}{HTML}{1B2430}

\resizebox{\columnwidth}{!}{%
\begin{tikzpicture}[
  font=\sffamily\footnotesize,
  every node/.append style={scale=.93},
  box/.style={rounded corners=2pt, line width=.6pt, minimum width=3.2cm, align=center, inner sep=1.5pt},
  llm/.style={box, draw=ovblue, fill=ovbluebg, text=ovblue, minimum height=.72cm},
  det/.style={box, draw=ovgreen, fill=ovgreenbg, text=ovgreen, minimum height=.72cm},
  agents/.style={llm, double copy shadow={shadow xshift=2pt, shadow yshift=2pt,
    draw=ovblue, fill=ovbluebg, line width=.45pt}},
  flow/.style={-{Latex[length=1.8mm,width=1.4mm]}, draw=ovdark, line width=.8pt},
  inner/.style={-{Latex[length=1.4mm]}, line width=.5pt},
  twoway/.style={inner, {Latex[length=1.4mm]}-{Latex[length=1.4mm]}, draw=ovblue},
  specflow/.style={-{Latex[length=1.5mm,width=1.2mm]}, draw=ovgray, dashed, line width=.6pt},
  edge/.style={font=\sffamily\scriptsize\bfseries, text=ovdark, align=left},
  innerlabel/.style={font=\sffamily\scriptsize, text=ovgray, align=center},
  note/.style={font=\sffamily\scriptsize\itshape, text=ovgray, align=center},
  badge/.style={circle, fill=ovdark, text=white, font=\sffamily\scriptsize\bfseries,
    minimum size=.42cm, inner sep=0pt},
]
\def\ovpanel#1#2#3#4#5#6#7{%
  \fill[ovpanel, draw=ovpanelline, line width=.5pt, rounded corners=4pt] (#1,#4) rectangle (#2,#3);
  \node[badge] at (#1+.35,#3-#7/2) {#5};
  \node[anchor=west, align=left, font=\sffamily\footnotesize\bfseries, text=ovdark]
    at (#1+.65,#3-#7/2) {#6};
}
\tikzset{spec/.style={draw=ovgray, dashed, line width=.7pt, rounded corners=2pt, fill=white,
  minimum width=3.0cm, minimum height=.75cm, align=center, font=\sffamily\footnotesize\bfseries}}

\ovpanel{0.1}{8.8}{-0.85}{-3.4}{1}{Generate search rules and construction plan}{0.5}
\ovpanel{0.1}{8.8}{-3.9}{-6.3}{2}{Find candidates}{0.5}
\ovpanel{0.1}{8.8}{-7.1}{-9.45}{3}{Construct instances}{0.5}
\ovpanel{0.1}{8.8}{-9.95}{-11.45}{4}{Package benchmark}{0.5}

\node[spec] (spec) at (4.45,0) {benchmark specification};
\node[box, draw=ovgray, fill=white, text=ovdark, minimum width=1.9cm, minimum height=.75cm,
  font=\sffamily\footnotesize\bfseries] (user) at (7.9,0) {researcher};
\draw[inner, draw=ovgray] (user.west) -- node[innerlabel, above]{writes,\\revises} (spec.east);

\node[llm] (a1) at (2.2,-1.85) {LLM writes\\search rules};
\node[llm] (j1) at (6.85,-1.85) {judge LLM\\reviews rules};
\draw[twoway] (a1.east) -- (j1.west);
\node[llm] (a3) at (2.2,-2.85) {LLM writes\\construction plan};
\node[llm] (j3) at (6.85,-2.85) {judge LLM\\reviews plan};
\draw[twoway] (a3.east) -- (j3.west);

\node[agents] (a2) at (2.2,-4.95) {search agents\\in parallel};
\node[note] at (2.2,-5.8) {$\circlearrowleft$ they exchange\\what they explore};
\node[det] (b2) at (6.85,-5.75) {existence check + dedup};
\draw[inner, draw=ovgreen] ([xshift=2pt]a2.east) -| node[innerlabel, above, pos=.45]{candidates + verification script} (b2.north);

\node[agents] (b3) at (2.2,-8.2) {construction agents\\in parallel};
\node[det] (c3) at (6.85,-8.2) {replay};
\draw[inner, draw=ovgreen] ([yshift=.15cm]b3.east) -- ([yshift=.15cm]c3.west);
\draw[inner, draw=ovgreen] ([yshift=-.15cm]c3.west) -- node[innerlabel, below]{replay\\report} ([yshift=-.15cm]b3.east);
\node[note] at (2.2,-9.0) {$\circlearrowleft$ they share lessons};

\node[det, minimum width=4cm] (a4) at (4.45,-10.95) {framework\\packages and checks};

\draw[flow] (spec.south) -- (4.45,-0.85);
\draw[flow] (1.0,-3.4) -- node[edge, right=2pt]{search rules} (1.0,-3.9);
\draw[flow] (b2.south) -- (6.85,-6.6) -- (1.0,-6.6) -- node[edge, right=2pt]{candidates} (1.0,-7.1);
\draw[flow] (a3.west) -- (-0.3,-2.85) -- (-0.3,-8.2) -- (b3.west);
\node[edge, rotate=90, anchor=south] at (-0.3,-5.5) {approved construction plan};
\draw[flow] (1.0,-9.45) -- node[edge, right=2pt]{instances + replay reports + construction plan} (1.0,-9.95);
\draw[flow] (1.0,-11.45) -- node[edge, right=2pt]{benchmark package} (1.0,-12.0);
\draw[specflow] (j1.east) -- (9.3,-1.85) -- (9.3,0) -- (user.east);
\draw[specflow, -] (j3.east) -- (9.3,-2.85) -- (9.3,-1.85);
\node[innerlabel, rotate=90, anchor=north] at (9.3,-1.42) {feedback if rejected};
\end{tikzpicture}}

\caption{Overview of the approach.}

\Description{A top-to-bottom diagram of four phase panels. A researcher writes the benchmark
specification, which enters Phase 1. In Phase 1, an LLM writes the search rules and a judge LLM
reviews them in a loop; then an LLM writes the construction plan and a judge LLM reviews it in a
loop. If a judge rejects the rules or the plan, its feedback returns to the researcher, who revises
the specification. The search rules pass to Phase 2, where several search agents that exchange what
they explore submit candidates with a verification script to an existence check and deduplication.
The candidates pass to Phase 3, and the approved construction plan passes from Phase 1 directly to
the construction agents of Phase 3, which work in parallel, share lessons, and receive replay
reports from replay. Instances, replay reports, and the construction plan pass to Phase 4, where the
framework packages and checks the instances and outputs the benchmark package.}
\label{fig:pipeline}
\end{figure}

\subsection{Problem Statement}
\label{sec:approach:problem}

BenchMeUp takes as input a natural-language \emph{benchmark specification} written by a researcher. The specification describes the task to be evaluated, the required contents of each instance, the sources from which instances may be obtained, and the conditions under which a technique's result is considered correct. The researcher also specifies the benchmark size. Table~\ref{tab:repair-spec} gives a simplified specification for program repair: each instance must provide a buggy Java program and a test that exposes the bug, and a repair technique must return a patch that makes the exposing test pass without introducing new test failures. We use program repair as a running example.

BenchMeUp produces a \emph{benchmark package}: a collection of instances and a common command-line interface through which evaluation code can use them. For program repair, an instance includes the versions and patches needed to reconstruct the buggy and fixed programs, an environment definition, and scripts to build and test them. Researchers can thus check out buggy versions, apply repair techniques, and invoke instance tests through a common interface.
The goal is to construct such a benchmark for any task within a given budget, without requiring researchers to implement the collection and execution of instances.

\subsection{Overview}
\label{sec:approach:overview}

Figure~\ref{fig:pipeline} gives an overview of how BenchMeUp constructs a benchmark package in four phases. In Phase~1, an LLM translates the researcher-written specification into shared \emph{search rules} that define where to search, which candidates to retain, and what information to record about them. In a separate query, an LLM generates, from the specification, a shared \emph{construction plan} that defines the artifacts to produce, the stages to follow, and the checks to perform. In Phase~2, \emph{search agents} use the search rules to search for and collect candidates from external data sources. For example, a candidate for program repair may consist of a repository, a fixing commit, and information about the changed program and test files.

In Phase~3, \emph{construction agents} follow the plan to construct instances from found candidates. After each stage of the plan, the \emph{framework} (our backend code that coordinates the agents) runs the agent-generated scripts in a fresh container and compares their results against the expected outcomes (defined in the plan). We call this procedure \emph{replay}. In Phase~4, the framework packages the completed instances behind a common command-line interface and verifies the packaged instances in fresh containers through this interface.

Because searching for candidates and constructing instances can take hours, BenchMeUp runs multiple agents in parallel. Search agents share summaries of their exploration to reduce overlap, while construction agents work on different candidates and share summaries of problems and attempted solutions. Algorithm~\ref{alg:main} presents the overall procedure, and Algorithm~\ref{alg:sync} describes agent scheduling and synchronization. We illustrate the four phases using the program-repair specification in Table~\ref{tab:repair-spec} and a candidate found for that specification.

\begin{algorithm}[t]
\caption{Constructing a benchmark from a specification.}
\label{alg:main}
\small
\renewcommand{\baselinestretch}{1.12}\selectfont
\begin{algorithmic}[1]
\Require specification $S$; targets $n_{\mathrm{search}}$, $n_{\mathrm{construction}}$
\Require agent counts $N_{\mathrm{search}}$, $N_{\mathrm{construction}}$
\PhaseBar{Phase 1: Generating Search Rules + Construction Plan}
\State feedback $F \gets \emptyset$ \label{l:rulesinit}
\Repeat
  \State $R \gets$ \Call{LLM-Generate-Or-Revise-Rules}{$S$, $F$} \label{l:rules}
  \State acceptance, $F \gets$ \Call{Judge}{$R$, $S$} \label{l:rulesjudge}
\Until{accepted or researcher-set iteration limit reached}
\If{the judge did not accept $R$}
  \State \Return $S$ and $F$ to the researcher \label{l:rulesreturn}
\EndIf \label{l:rulesend}
\State feedback $F \gets \emptyset$ \label{l:planinit}
\Repeat
  \State $P \gets$ \Call{LLM-Generate-Or-Revise-Plan}{$S$, $F$} \label{l:plan}
  \State acceptance, $F \gets$ \Call{Judge}{$P$, $S$, $R$}
\Until{accepted, iteration limit, or missing researcher input}
\If{the judge did not accept $P$}
  \State \Return issue to researcher; request missing input if needed \label{l:planreturn}
\EndIf \label{l:planend}
\PhaseBar{Phase 2: Finding Candidates}
\State $C \gets \emptyset$\Comment{recorded candidates}
\State $\mathit{agents}_{\mathrm{s}} \gets$ \Call{Start-Search-Agents}{$S$, $R$, $N_{\mathrm{search}}$, $C$} \label{l:searchstart}
\State \textbf{every $\mathit{synctime}_{\mathrm{s}}$:} \Call{Synchronize}{$\mathit{agents}_{\mathrm{s}}$}\Comment{Algorithm~\ref{alg:sync}} \label{l:searchsync}
\State stop search when $|C| \ge n_{\mathrm{search}}$ or search time expires
\State $C \gets$ \textsc{Deduplicate}($C$) \label{l:merge}
\PhaseBar{Phase 3: Turning Candidates into Instances}
\State $I \gets \emptyset$, $L \gets \emptyset$ \Comment{instances, lessons}
\State $\mathit{agents}_{\mathrm{c}} \gets$ start $N_{\mathrm{construction}}$ agents on $(S, P, C, L)$ \label{l:instantiate}
\State \textbf{every $\mathit{synctime}_{\mathrm{c}}$:} $L \gets$ \Call{Synchronize}{$\mathit{agents}_{\mathrm{c}}$} \label{l:consync}
  \ForAll{agent \textbf{in} $\mathit{agents}_{\mathrm{c}}$, \textbf{in parallel}} \label{l:agents}
    \While{$|I| < n_{\mathrm{construction}}$ and budget remains}
      \If{no candidates remain}
        \State stop construction
      \EndIf
    \State $c \gets$ next candidate from $C$; artifacts $A_c \gets \emptyset$
    \ForAll{stage $s$ of $P$} \label{l:stage}
      \Repeat
        \State agent creates or revises files in $A_c$ following $s$
        \State \Call{Replay}{$s$, $A_c$} in a fresh container \label{l:replay}
      \Until{checks pass, attempt limit, or agent drops $c$}
      \If{checks do not pass}
        \State drop $c$; \textbf{break} \Comment{next candidate}
      \EndIf
    \EndFor
    \If{not dropped}
      \State add instance $(c, A_c)$ to $I$ \label{l:add}
    \EndIf
    \EndWhile
  \EndFor
\PhaseBar{Phase 4: Packaging the Benchmark}
\State $benchmark \gets$ package each instance in $I$ with replay and interface \label{l:packagestart}
\State \Return $benchmark$ \label{l:package}
\end{algorithmic}
\end{algorithm}

\begin{table}[t]
\centering
\caption{Researcher-written benchmark specification for program repair.}
\label{tab:repair-spec}
\footnotesize
\begin{tabular}{@{}lL{0.8\columnwidth}@{}}
\toprule
\textbf{Part} & \textbf{Text} \\
\midrule
Task & Repair a real bug in a Java project. \\
Instance & Provide a buggy program and a test exposing the bug. The repair technique returns a source-code patch. \\
Data source & Real, public Java projects that are popular, actively maintained, have substantial development histories, and contain resolved bug reports. \\
Correctness & The exposing test fails before the fix and passes after applying the submitted patch. Previously passing tests must not start failing. \\
Constraints & Exclude projects already contained in Defects4J. \\
\bottomrule
\end{tabular}
\end{table}

\subsection{Phase 1: Generating Search Rules and Construction Plan}
\label{sec:approach:formulation}

Phase~1 derives, from the specification, the two artifacts that the agents of the later phases use: the search rules for Phase~2 and the construction plan for Phase~3.

\subsubsection{Generating the Search Rules}

A specification describes the intended benchmark, but rarely settles every decision needed during search. In Table~\ref{tab:repair-spec}, for example, ``popular'' and ``actively maintained'' do not give numerical thresholds, while ``real bug'' requires distinguishing behavior-changing fixes from style or feature changes. If search agents fill in these details independently, one might accept a repository with 100 stars while another agent requires 1,000 stars; they might also disagree about which commits count as fixes. The found candidates would then reflect different selection policies even though they belong to the same benchmark.

Phase~1 makes those choices once, before the search agents begin. The framework prompts an LLM to fill a task-independent search-rule schema from the specification. The resulting rules define how to identify a candidate, where agents may look for it, which conditions allow keeping it, and what information to record. If the specification does not mention a data source, the LLM selects one (e.g., GitHub). The rules are saved as YAML so they can be inspected and reused. They give agents common criteria, although agents must still assess whether individual candidates meet those criteria.

\paragraph{Running example.} For the repair specification, the generated rules in Table~\ref{tab:repair-rules} identify a candidate by repository URL and fixing commit. They operationalize ``popular'' as at least 100 stars, require substantial development history and recent activity, and retain commits that appear to fix bugs while changing both program and test code. A fixing commit that changes no tests may be hard to turn into the specified fail-to-pass repair instance. The recorded commit message and changed-file information give Phase~3 a good starting point.

\begin{table}[t]
\centering
\caption{Search rules generated for program repair (excerpt).}
\label{tab:repair-rules}
\footnotesize
\setlength{\tabcolsep}{1.5pt}
\begin{tabular}{@{}lL{0.75\columnwidth}@{}}
\toprule
\textbf{Rule} & \textbf{Value} \\
\midrule
Candidate identity & Repository URL, fixing commit \\
Source & Java; at least 100 stars (popular), 1,000 commits (substantial history), and 20 closed bug reports (resolved bug reports); active within the last 12 months (actively maintained) \\
Retain when & The commit appears to fix a real bug; program code is changed; test code is added or modified \\
Record & Commit message, changed program file, changed test file \\
\bottomrule
\end{tabular}
\end{table}

Before using the rules, a second LLM judges whether the proposed rules omit requirements or leave important ambiguities unresolved. If the judge rejects the rules, it returns the feedback to the generating LLM, which revises them for another review. The framework repeats this loop until the judge accepts the rules or a researcher-set iteration limit is reached. If the limit is reached, the specification and feedback are returned to the researcher rather than starting a search with unapproved rules (Algorithm~\ref{alg:main}, lines~\ref{l:rulesinit}--\ref{l:rulesend}). Judge approval only checks the formulation against the specification; it does not establish that the resulting criteria are objectively correct.

\subsubsection{Generating the Construction Plan}
After the judge accepts the search rules, the framework asks an LLM to generate the \emph{construction plan} from the benchmark specification. The plan describes how to turn a candidate into a benchmark instance, including the steps to follow, the artifacts to produce, and the checks needed to verify the instance. It divides construction into stages, each with an objective and required artifacts. All construction agents follow the same plan, but each determines how to carry out its steps for the particular candidate it is working on.

\paragraph{Running example.} Listing~\ref{lst:repair-plan} shows an excerpt of the construction plan generated for program repair. The plan guides construction agents through a sequence of stages, starting with setting up the repository and its environment, then separating the fixing commit into program and test patches, building the buggy and fixed versions, and creating scripts to run the required tests. It also specifies the outcomes that must be demonstrated. For example, the exposing tests fail on the buggy version and pass on the fixed version, and removing the known fix restores the failure. In addition, the plan defines common commands for checking out versions, applying patches, building, and running tests. These commands are later exposed through the benchmark's interface, allowing techniques to be evaluated consistently across instances.

\lstdefinestyle{plan}{escapeinside={(*}{*)},
  emph=[1]{commands,oracle,fix_removal_check,technique_input,stages,name,files,done_when},
  emph=[2]{}, emph=[3]{}, emph=[4]{}}
\begin{lstlisting}[float=htbp, style=plan,
  caption={Construction plan generated for program repair (excerpt, paraphrased). ``...'' marks omitted text.},
  label={lst:repair-plan}]
commands:                   # interface commands
  checkout <dir> <commit>   # working tree at <commit>
  apply_patch <dir> <patch> # apply the patch with git
  build <dir>               # compile, run no tests
  run_tests <dir>           # exposing tests, JUnit XML
  check_all <dir>           # full suite, no new failures
  revert_fix <dir>          # remove the fix, rebuild
(*\color{listingkey}\bfseries versions*):
  buggy: [checkout {buggy_commit},
          apply_patch test.patch, build]
  fixed: [checkout {fixing_commit}, build]
oracle: {run_tests fails on buggy, passes on fixed}
fix_removal_check: {revert_fix on fixed: tests fail}
technique_input: {apply_patch on buggy: accepts fix.patch,
                  rejects empty.patch}
stages:
  - name: identify the bug, install the JDK, and clone
    files: [Dockerfile, setup.sh, instance.json]
    done_when: the image builds, setup.sh exits 0,
      instance.json records the exposing tests and the
      buggy commit (the parent of the fixing commit), ...
  - name: split the fixing commit, build both versions
    files: [checkout.sh, apply_patch.sh, build.sh,
            fix.patch, test.patch, empty.patch]
    done_when: both versions build in a fresh container; ...
  - name: write run_tests.sh, check_all.sh, revert_fix.sh
    done_when: oracle and fix_removal_check hold, and
      check_all.sh exits 0 on both versions
  - name: evaluate fix.patch and empty.patch
    done_when: technique_input holds
\end{lstlisting}

The plan also goes through a review step. A second LLM judges the proposed plan against the specification and search rules, returning feedback when revisions are needed. The generating LLM revises and resubmits the plan until it is accepted or the feedback-iteration limit is reached. If the judge asks for information missing from both the specification and rules and the generator LLM cannot supply it, the framework asks the researcher to supplement the missing information. If the limit is reached, the framework returns the unapproved plan and the feedback to the researcher (Algorithm~\ref{alg:main}, lines~\ref{l:planinit}--\ref{l:planend}). The review process helps identify missing requirements and ambiguities in the construction plan before construction agents begin working on candidates.

\subsection{Phase 2: Finding Candidates}
\label{sec:approach:search}

Phase~2 uses the specification and the accepted search rules to assemble a pool of candidate records. The goal is to find plausible instances satisfying the search rules; the costlier creation of runnable instances waits until Phase~3.

\subsubsection{Searching with Agents}

Search agents autonomously explore external data sources to find candidates that satisfy the search rules. Rather than following a predefined sequence of queries, each agent decides how to proceed based on the information it obtains during the search. It can write and execute scripts to query data sources, inspect the returned records, and investigate promising results further. For each potential candidate, the agent checks available information against the search rules' selection conditions and records the required fields if suitable. The agent then continues exploring the data source to find additional candidates.

\paragraph{Running example.} When searching for program-repair instances, an agent may first query GitHub for Java repositories that meet the requirements on popularity, development history, and activity. It can then inspect their commit histories to identify potential bug fixes and examine the corresponding changes to program and test files. For commits that satisfy the selection conditions, it records the repository URL, fixing commit, and changed files, as specified in Table~\ref{tab:repair-rules}. One recorded candidate, used in the following phases, is commit \code{7d7043d} of \code{52North/SOS} (``Fix bug with observation identifier becoming too long''), together with its changed program file and test file.

Search agents use a customized agent scaffold based on mini-swe-agent~\cite{yang2024sweagent} and inspired by agents such as ExecutionAgent~\cite{bouzenia2025you}. Each search agent receives the original benchmark specification, search rules, and task-independent instructions on its tools, search procedure, and output format.

\subsubsection{Coordinating Search Agents}

A single search agent may explore only a small portion of the available data sources before its time budget expires. BenchMeUp therefore runs multiple search agents in parallel. However, agents working independently may issue similar queries and repeatedly investigate the same repositories or candidates. To reduce this overlap, the framework starts agents at fixed intervals and shares information about their search activities.

The first agent starts alone. Before the next agent starts, an LLM summarizes the first agent's search directions, including its queries, filters, and explored repositories. The next agent receives this summary and is instructed to explore different directions while following the same search rules. Each subsequent agent similarly receives summaries of preceding agents' exploration. Once all $N_{\mathrm{search}}$ agents are running, the framework periodically updates and distributes these summaries at intervals of $\mathit{synctime}_{\mathrm{s}}$, allowing agents to adjust their searches based on the exploration of others.

The researcher sets the candidate target $n_{\mathrm{search}}$, search-agent count, start delay, synchronization interval, and search time budget. The number of agents is also constrained by available computing resources and service limits, since agents querying GitHub or Hugging Face, or using Docker, compete for shared resources. Algorithm~\ref{alg:main} (lines~\ref{l:searchstart} and~\ref{l:searchsync}) and Algorithm~\ref{alg:sync} describe this coordination procedure.
\begin{algorithm}[t]
\caption{Synchronization of parallel agents.}
\label{alg:sync}
\small
\renewcommand{\baselinestretch}{1.12}\selectfont
\begin{algorithmic}[1]
\Function{Start-Search-Agents}{$S$, $R$, $N_{\mathrm{search}}$, $C$}
  \State agents add candidates to $C$ after existence checks pass
  \State $\mathit{agents} \gets \{$agent 1 started on $(S, R)\}$
  \For{agent $j = 2$ \textbf{to} $N_{\mathrm{search}}$}
    \State wait for the start delay \label{l:stagger}
    \State summaries $\gets$ \Call{Summarize}{trajectories of $\mathit{agents}$}
    \State start agent $j$ on $(S, R)$ and summaries; add it to $\mathit{agents}$
  \EndFor
  \State \Return $\mathit{agents}$
\EndFunction
\Statex
\Function{Synchronize}{$\mathit{agents}$}
  \ForAll{agent \textbf{in} $\mathit{agents}$, \textbf{in parallel}}
    \State summary of agent $\gets$ \Call{Summarize}{trajectory of agent}\label{l:summ}
  \EndFor
  \If{$\mathit{agents}$ are search agents} \label{l:ifsearch}
    \State send each agent the summaries of all other agents \label{l:searchshare}
  \Else
    \State lessons $L \gets$ \Call{Merge}{summaries of all agents}
    \State send $L$ to each agent \label{l:conshare}
    \State \Return $L$
  \EndIf
\EndFunction
\end{algorithmic}
\end{algorithm}

\subsubsection{Checking and Deduplicating Candidates}

Before a candidate joins the shared pool, its search agent must provide an existence-verification script. Each agent writes one such script for all of its candidates; scripts can differ because agents may use different data sources. The framework executes the script in a fresh container and checks recorded identities and fields against the source, detecting records such as nonexistent commit hashes. Only candidates that pass are recorded.

Phase~2 ends when the shared pool reaches $n_{\mathrm{search}}$ candidates or the search time budget expires. The framework deduplicates the recorded pool before passing candidates to Phase~3 (Algorithm~\ref{alg:main}, line~\ref{l:merge}).

\subsection{Phase 3: Turning Candidates into Instances}
\label{sec:approach:construction}

A candidate record identifies promising source material, but it is not yet something a researcher can use to evaluate a technique. A bug-fixing commit, for example, contains a program fix and a test change. To make it a repair instance, BenchMeUp must reconstruct a buggy program that includes the exposing test but excludes the fix, provide a build environment and scripts, show that the test fails before repair and passes afterward, and provide a way to judge a submitted patch. Phase~3 performs this transformation for each candidate, producing instance-specific artifacts $A_c$ and adding successfully constructed instances to a collection $I$.

Construction agents follow the construction plan from Phase~1 to prepare instances from individual candidates. Since an instance may work in an agent's environment but fail elsewhere, the framework replays the generated scripts in fresh containers to check that they produce the expected results. 

\subsubsection{Constructing and Replaying Instances}

Each construction agent takes a candidate $c$ and works through the plan's stages in order. At each stage, the agent prepares the required artifacts, such as instance-specific metadata, environment configurations, and scripts (referred to as $A_c$ in the algorithm), and then submits them to the framework for checking.

Rather than relying on the agent's reported results, the framework copies the instance files into a fresh container and runs the setup and checks specified by the plan. It compares the observed results with the expected outcomes and reports any failures to the construction agent. The agent can then revise its artifacts and submit the stage again. The framework also reruns checks from earlier stages to detect regressions caused by subsequent changes. The agent can proceed to the next stage only after all required checks of the current stage pass.

\paragraph{Running example.} For the SOS candidate, the construction agent identifies the added test as the exposing test and separates the fixing commit into \code{test.patch}, containing the test and changes needed to compile it, and \code{fix.patch}, containing the program repair. This allows the agent to prepare a buggy version that includes the exposing test without the fix. Replay then checks that the original tests pass on the parent commit, the exposing test fails on the buggy version, and the same test passes on the fixed version. The framework also checks that the generated scripts correctly evaluate the known fix and reject an empty patch.

If a required check continues to fail after a researcher-set number of attempts, the framework drops the candidate and assigns the agent another candidate. A construction agent can also drop a candidate when it finds that the candidate does not satisfy the instance criteria; it then documents its evidence. If all stages pass, the framework adds $(c,A_c)$ to $I$ for packaging (Algorithm~\ref{alg:main}, lines~\ref{l:stage}--\ref{l:add}). The researcher sets a construction target $n_{\mathrm{construction}}$ and time budget. Construction stops at the target, time limit, or candidate exhaustion.

\subsubsection{Coordinating Construction Agents}

Constructing an instance can take considerable time because agents may need several attempts to configure dependencies, build projects, and pass replay checks. BenchMeUp therefore runs $N_{\mathrm{construction}}$ construction agents in parallel, each working on a different candidate. Although the agents work on separate candidates, they may encounter similar problems, such as missing dependencies or build failures. To help agents learn from one another's experience, the framework periodically uses an LLM to summarize their trajectories, including problems encountered, attempted solutions, and replay feedback. It shares these summaries with the construction agents as lessons $L$, helping them reuse successful solutions and avoid previously unsuccessful attempts. The researcher sets the synchronization interval $\mathit{synctime}_{\mathrm{c}}$.

An agent's context grows with every step it takes, which makes its later model calls more expensive and its mistakes more likely. The framework therefore runs construction agents in waves. Each agent works for a fixed time window, during which it can construct several candidates one after another. When the window ends, the framework replaces the agent with one prompted with the current lessons $L$ and assigns it the next candidates. An agent that is in the last stage of an instance when its window ends gets a grace period to finish it. The researcher sets the window length and the grace period.

Shared experience relates to prior agent-learning approaches~\cite{shinn2023reflexion,zhao2024expel,wang2025awm}; our summaries include replay feedback from executing the agents' scripts alongside the agents' own accounts. Algorithm~\ref{alg:main} (lines~\ref{l:instantiate}--\ref{l:consync}) and Algorithm~\ref{alg:sync} describe construction coordination.

\subsection{Phase 4: Packaging the Benchmark}
\label{sec:approach:packaging}

After Phase~3, each completed instance has its own files and scripts, created for its particular repository. Those scripts may satisfy the shared plan, but a researcher still needs a uniform way to invoke them. Phase~4 turns the collection $I$ into a benchmark package whose public commands have the meanings defined by the construction plan, regardless of which agent produced the underlying scripts.

For each completed instance, the framework copies $A_c$ and retains the replay reports so users can inspect the checks performed during construction. It also writes a manifest mapping the plan's common commands to that instance's scripts and files. A shared command-line program reads the manifest and dispatches requests to the selected instance. Evaluation code can consequently obtain instance files, prepare a working tree, invoke tests, or perform other task-specific operations without knowing each instance's internal script layout. Packaging can start upon instance delivery, alongside ongoing construction.

\paragraph{Running example.} Listing~\ref{lst:repair-interface} illustrates the evaluation workflow for SOS: check out the buggy parent, apply the exposing-test patch, build, apply the technique's patch, and invoke the exposing tests and suite-wide checks. The scripts implementing these steps are specific to SOS, but the commands used to invoke them are those defined for the repair benchmark as a whole. Phase~4 enforces this consistency.

Finally, the framework reruns the packaged interface in a fresh container built from the instance's Dockerfile. This differs from staged replay in Phase~3: it checks the completed instance as exposed through the package, not just the intermediate artifacts submitted during construction. Instances whose packaged commands fail the plan's required checks are excluded. Algorithm~\ref{alg:main}, lines~\ref{l:packagestart}--\ref{l:package}, specifies this final step.

\begin{lstlisting}[float=t,
caption={Invoking the command-line interface on an instance of the program-repair benchmark (illustrative command syntax).},
label={lst:repair-interface}]
benchmark --instance SOS-7d7043d checkout <dir> {buggy_commit}
benchmark --instance SOS-7d7043d apply_patch <dir> test.patch
benchmark --instance SOS-7d7043d build <dir>
benchmark --instance SOS-7d7043d apply_patch <dir> <submitted.patch>
benchmark --instance SOS-7d7043d run_tests <dir>
# the exposing tests must pass
benchmark --instance SOS-7d7043d check_all <dir>
# must exit 0: no new test failures
\end{lstlisting}

\section{Evaluation}
\label{sec:eval}

We organize the evaluation around four research questions:

\begin{enumerate}
\item[RQ1] How effectively and efficiently does BenchMeUp construct benchmarks for software-engineering tasks?
\item[RQ2] To what extent do constructed benchmarks support end-to-end evaluation of software-engineering techniques?
\item[RQ3] What are our benchmarks' characteristics and quality?
\item[RQ4] How does BenchMeUp compare with a general-purpose LLM agent?
\end{enumerate}

\subsection{Experimental Setup}
\label{sec:setup}

\subsubsection{Tasks}
\begin{table*}[t]
\centering
\caption{Tasks and construction outcomes. Minutes (Min.) and USD are per constructed instance.}
\label{tab:tasks}
\footnotesize
\setlength{\tabcolsep}{3.5pt}
\begin{tabular}{@{}>{\raggedright\arraybackslash}p{0.2\textwidth}>{\raggedright\arraybackslash}p{0.23\textwidth}rrrrrrrrr@{}}
\toprule
 & & \multicolumn{1}{c}{Spec} & \multicolumn{3}{c}{Search} & \multicolumn{4}{c}{Construction} & Packaging \\
\cmidrule(lr){3-3}\cmidrule(lr){4-6}\cmidrule(lr){7-10}\cmidrule(l){11-11}
One benchmark instance & Acceptance condition & Words & Found & Min. & USD & Attempted & Constructed & Min. & USD & Packaged \\
\midrule
\multicolumn{11}{@{}l}{\textbf{Program repair}} \\
A bug-fixing commit with a bug-exposing test & The test fails before the fix and passes after it & 1,118 & 176 & 1.9 & 0.47 & 106 & 50 & 42.8 & 0.30 & 50 \\
\addlinespace[2pt]
\midrule
\multicolumn{11}{@{}l}{\textbf{Issue reproduction}} \\
An issue-closing pull request that adds or changes a test & The test fails before the pull request and passes after it & 858 & 200+ & 1.4 & 0.35 & 88 & 50 & 19.3 & 0.25 & 49 \\
\addlinespace[2pt]
\midrule
\multicolumn{11}{@{}l}{\textbf{Compiler autotuning}} \\
A runnable C program at a repository commit & Candidate and reference builds produce the same output & 2,381 & 60 & 2.9 & 1.08 & 52 & 50 & 11.9 & 0.18 & 50 \\
\addlinespace[2pt]
\midrule
\multicolumn{11}{@{}l}{\textbf{Fixing static-analysis warnings}} \\
A SonarQube warning at a specific location + fix & The target warning disappears without introducing a new finding & 805 & 197 & 3.5 & 0.12 & 86 & 48 & 60.0 & 0.40 & 47 \\
\midrule
\multicolumn{2}{@{}l}{\textbf{Average}} & 1,291 & 160 & 2.4 & 0.51 & 83 & 49.5 & 33.5 & 0.28 & 49 \\
\bottomrule
\end{tabular}

\smallskip
\end{table*}

We evaluate the approach on the four tasks shown in Table~\ref{tab:tasks}. We select these tasks for three reasons: i) they are common software-engineering tasks with existing benchmarks that we can compare to; ii) they cover different languages (Java, Python, C) and different kinds of instances; and iii) the tasks impose different construction requirements, including commit inspection, dependency setup, test execution, static analysis, and runtime measurements. For each task, we wrote an initial benchmark specification and revised it based on the feedback from BenchMeUp's LLM judges (0--6 revisions per task). The table reports final specification word counts.

\subsubsection{Models and Budgets}
We use deepseek-v4-flash for construction by default and Claude Opus~5 for search, rules, and plan generation. Construction requires extensive interaction with repositories and their environments, motivating the lower-cost model. Section~\ref{sec:discussion} discusses construction-model effects.

To explore and construct in parallel, each task uses four search agents, each limited to six hours, and four construction agents with a wall-clock budget of 48 hours (192 agent-hours). Search agents synchronize every 20 minutes and construction agents every 60 minutes; search agents start 20 minutes apart. All experiments run on one machine with 32 virtual cores and 62.8\,GB of memory (Ubuntu 24.04). Compiler-autotuning measurements run sequentially.

\subsection{RQ1: Effectiveness and Efficiency of BenchMeUp}
\label{sec:rq1}

For each of the four tasks, we set a search target of 200 candidates and a construction target of 50 instances. To avoid waiting for the search to finish, BenchMeUp launches construction agents as soon as at least 50 candidates have been collected. Search and construction then proceed concurrently: search agents continue finding candidates while construction agents turn the available candidates into instances. Once 50 instances have been constructed for a task, the framework stops both phases, even if the search agents have not reached their target of 200 candidates. This stopping rule is a choice of our experiments; a researcher can let both phases continue.

\subsubsection{Effectiveness}
Table~\ref{tab:tasks} summarizes the effectiveness of benchmark construction. Across the four tasks, search agents found 639 unique candidates. Construction agents tried 332 candidates, constructing 198 instances and dropping 134.
The program-repair, issue-reproduction, and compiler-autotuning benchmarks meet the target of 50 instances; the static-analysis benchmark reaches only 48 constructed instances within the time limit. Of the 198 constructed instances, 196 (99\%) pass the final packaging checks.

The 134 candidates dropped during construction fall into two categories. First, the construction agent decides that a candidate does not satisfy the instance criteria, contrary to the search agent's initial assessment, and documents the supporting evidence. Second, the construction agent is unable to construct a valid instance from the candidate. The second category reflects, at least in part, limits of the construction model. As we discuss in Section~\ref{sec:discussion}, a stronger construction model completes construction faster and drops fewer candidates.

During construction, replay helps to identify instances that fail outside of the agent's container. Of the 198 constructed instances, 108 (55\%) initially had at least one failed replay, totaling 170 failures. Most failures are due to builds relying on residual container state (82) or test scripts failing in the replay environment (47). BenchMeUp returns these failures to the construction agents for correction and accepts the instances only after successful replay in a fresh container. However, replay does not establish whether instance checks distinguish correct from incorrect inputs; RQ3 evaluates this.

\subsubsection{Time and Cost}
Table~\ref{tab:tasks} also reports the time and LLM cost, focusing on search and construction, as these are the most expensive steps. Across the four tasks, the average LLM cost is \$0.79 per constructed instance, of which 64\% is for search and 36\% for construction. Search costs more because it uses the stronger model. Looking at time, the proportions reverse: search accounts for only 4--20\% and construction for 80--96\% of the total time. On average, an instance is found and constructed in 36 minutes.

\subsection{RQ2: Existing Techniques on the Constructed Benchmarks}
\label{sec:rq3}

To assess the usefulness of the constructed benchmarks, we run existing techniques that were previously developed for the corresponding tasks on the packaged instances.
For program repair, we use RepairAgent~\cite{bouzenia2025repairagent} with GPT-5-mini and GPT-5-nano, using the original technique's default parameters. For issue reproduction, we run ReProAgent~\cite{zhang2026reproagent} and mini-swe-agent~\cite{yang2024sweagent} (version 2.4.6), both with GPT-5-mini. For compiler autotuning, we adapt BOCA~\cite{chen2021boca} to reduce compiled-program runtime. For fixing SonarQube static-analysis warnings, we evaluate google-java-format~\cite{googlejavaformat}, Sorald~\cite{etemadi2022sorald}, OpenRewrite~\cite{openrewrite}, and CodeCureAgent~\cite{joos2026codecureagent}.

\begin{figure}[t]
\centering
\includegraphics[width=\columnwidth]{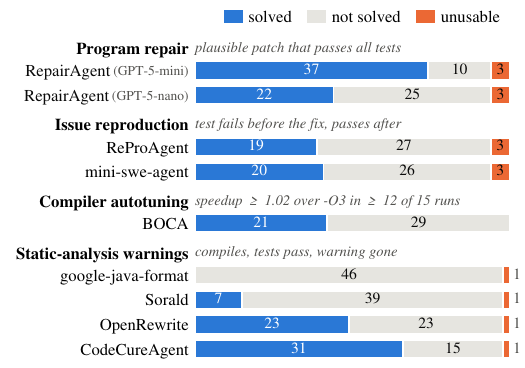}
\caption{Outcomes of the existing techniques on the packaged instances of each task: instances the technique solves, does not solve, or cannot use.}
\label{fig:technique-results}
\end{figure}

Figure~\ref{fig:technique-results} reports, for each technique, how many instances it solves, fails to solve, or cannot run on (``unusable'').
Of the 196 packaged instances, 189 (96\%) can be used successfully by the evaluated techniques.
In RQ3, we give more details about the seven unusable instances. 

The different techniques solve different numbers of instances.
RepairAgent produces plausible patches for 37 of the 47 usable program-repair instances with GPT-5-mini and for 22 with GPT-5-nano. ReProAgent and mini-swe-agent reproduce 19 and 20 of the 46 usable issues, respectively. BOCA finds configurations that meet our speedup criterion for 21 of the 50 programs. For fixing static-analysis warnings, CodeCureAgent fixes the most instances, followed by OpenRewrite. This order matches what the techniques are designed for: google-java-format only reformats code, Sorald and OpenRewrite apply transformations for the rules they support, and CodeCureAgent uses an LLM to address any rule.
In program repair, the developer fixes of the instances that RepairAgent with GPT-5-mini does not solve have a median of 25 changed lines, compared with eight for the solved ones. For compiler autotuning, BOCA runs on all 50 programs, but no explored configuration consistently beats \code{-O3} by 2\% on unsolved programs. For fixing static-analysis warnings, CodeCureAgent validates its patches with an older SonarQube version (9.9 vs 25.x), thus missing some rules.

\subsection{RQ3: Characteristics and Quality of Benchmarks}
\label{sec:quality}

We first characterize all constructed instances quantitatively and compare them with established benchmarks of the same tasks. We then inspect a sample of the instances qualitatively.

\subsubsection{Quantitative Characterization}
Table~\ref{tab:charac} lists the measures. For program repair, issue reproduction, and fixing static-analysis warnings, we examine the number of projects, their popularity measured by GitHub stars, and the distribution of instances across projects. We also measure the patch size, as in prior analyses of Defects4J~\cite{sobreira2018dissection} and SWE-bench~\cite{Rondon2025}. Finally, we include task-specific measures, such as the proportion of single-file fixes in program repair and program size, build time, and workload count in compiler autotuning.

\begin{table}[t]
\centering
\caption{Characteristics of the instances used for evaluation compared to existing benchmarks of the same task.}
\label{tab:charac}
\footnotesize
\setlength{\tabcolsep}{5pt}
\begin{tabular}{@{}lrr@{}}
\toprule
Measure & Ours & Reference \\
\midrule
\textbf{Program repair} & & Defects4J~\cite{just2014defects4j} \\
Projects (count) & 31 & 17 \\
GitHub stars (median) & 141 & 1,391 \\
Instances per project (mean) & 1.6 & 49 \\
Patch size (median lines) & 13 & 6 \\
One-file fixes (share) & 82\% & -- \\
\midrule
\textbf{Issue reproduction} & & SWT-bench Verified~\cite{mundler2024swtbench} \\
Projects (count) & 33 & 12 \\
GitHub stars (median) & 199 & 14,738 \\
Instances per project (mean) & 1.5 & 36 \\
Patch size (median lines) & 25 & 6 \\
Added test lines (median) & 41.5 & 14 \\
Pull requests merged in 2026 (count) & 37 & 0 \\
\midrule
\textbf{Compiler autotuning} & & cBench~\cite{fursin2009collective} \\
Lines of C code per project (median) & 14,700 & 1,780 \\
C files per project (median) & 19 & 6 \\
Workloads per program (median) & 4 & 20 \\
Build time at \texttt{-O3} (s) & 2.40 & 0.45 \\
Runtime at \texttt{-O3} (median, s) & 1.39 & 1.94 \\
Runtime at \texttt{-O0} / at \texttt{-O3} (median) & 1.73 & 1.84 \\
\midrule
\textbf{Fixing static-analysis warnings} & & Baseline~\cite{joos2026codecureagent} \\
Projects (count) & 18 & 106 \\
GitHub stars (median) & 9,252 & 260 \\
Patch size (median lines) & 10 & -- \\
Instances per rule (mean) & 1.95 & 3.43 \\
Build tool: Maven/Ant/Gradle (count) & 20/18/9 & 1,000/0/0 \\
\bottomrule
\end{tabular}
\end{table}

\paragraph{Projects.}
The constructed program-repair and issue-reproduction benchmarks cover more projects than their reference benchmarks, although these projects generally have fewer GitHub stars. None of the 31 program-repair projects appears in Defects4J or GitBug-Java~\cite{silva2024gitbugjava}, extending coverage to projects absent from these benchmarks. Project popularity depends partly on the minimum-star thresholds in the generated search rules (Table~\ref{tab:repair-rules}), which researchers can adjust to target more popular projects.

\paragraph{Recency.}
Most issue-reproduction instances come from pull requests merged in 2026. Such recent instances may be less likely to appear in the evaluated models' training data, although we do not measure contamination.

\paragraph{Size of programs and patches.}
The constructed benchmarks contain larger patches than Defects4J and SWT-bench, and larger programs than cBench~\cite{fursin2009collective}. None of the program-repair fixes changes only one line. Larger changes may pose greater challenges for existing techniques. For example, RepairAgent fails more often on instances with larger patches (Section~\ref{sec:rq3}). Finally, unlike the dataset used to evaluate CodeCureAgent~\cite{joos2026codecureagent}, our benchmark also includes developer fixes verified by rerunning SonarQube.

\subsubsection{Qualitative Inspection}
Running existing techniques on our instances provides evidence of their usability, but does not establish their quality. For example, do the program-repair instances represent genuine bug fixes, or do they contain unrelated changes? Answering such questions requires examining individual instances in the context of their tasks. We therefore manually inspect a sample of 40 instances.
For each task, we randomly select five instances solved by the best-performing technique from RQ2 and five instances it does not solve. We first check each instance against its upstream repository, verifying the specification criteria. We test interfaces with recorded inputs, such as the developer fix, and incorrect ones, e.g., random patches or empty submissions.

Of the 40 inspected instances, 37 are valid instances of their tasks. The three exceptions belong to issue reproduction. Two instances have an incorrect problem description because the construction agent wrote its own description rather than using the original GitHub issue text. In the third instance, the reference test exercises a helper unrelated to the reported issue. We also find minor problems, such as bug fixes containing unrelated formatting changes and issues describing feature requests, which also occur in SWE-bench.

Beyond this sample, seven packaged instances cannot be used by the techniques in RQ2: three in program repair, three in issue reproduction, and one in fixing static-analysis warnings (Figure~\ref{fig:technique-results}). The causes include changes to external dependencies after construction, such as removed build files or incompatible pip versions.

Finally, each instance provides task-specific basic commands, such as applying a patch, building, or running tests, which the techniques in RQ2 use. In all 40 inspected instances, the basic commands handle our incorrect inputs correctly. In six instances, however, the joint ``check'' command, which combines the basic commands to decide whether a submission is correct, misses special cases. For example, a patch that modifies a test to mark it as skipped leaves no failing test, so the check command accepts the patch, whereas the basic command for running tests reports that no test ran. These cases do not affect RQ2 (relies on basic commands), but researchers using the joint command need to check such details.

\subsection{RQ4: Comparison with a General-Purpose Agent}

As a baseline, we give Codex~\cite{openai2025codex} (CLI 0.163.0) the specification of each task and ask it to construct the benchmark with the same model as our construction agents. We launch Codex sessions in parallel, each in its own workspace. We ask each session for five instances, because larger requests make Codex stop early or exhaust its context window. A session is limited to 8 hours and \$2.50, which no session reached.

We assess every instance of both approaches against seven successive criteria (Table~\ref{tab:codex-gates}). The criteria follow an instance from its source to its evaluation. The first three criteria are about selection: an instance must come from a real source (real), satisfy the specification's filters (admissible), and be unique. The next two are about reproducibility: an instance must contain all required files (complete) and have a buildable environment (builds). The last two concern the evaluation interface: its commands must run (functional) and produce the expected outcome (expected). These three groups correspond to the three challenges of Section~\ref{sec:intro}. The order follows the dependencies between the criteria. For example, building an instance is meaningless if its source does not exist. We automatically check all criteria using scripts we created, manually tested, and adapted to each task's definition and specification.

\begin{table}[t]
\centering
\caption{Instances delivered by BenchMeUp and by Codex that meet each successive criterion.}
\label{tab:codex-gates}
\footnotesize
\setlength{\tabcolsep}{2pt}
\newcommand{\twoline}[2]{\scriptsize\begin{tabular}[b]{@{}c@{}}#1\\#2\end{tabular}}
\begin{tabular*}{\columnwidth}{@{}l@{\extracolsep{\fill}}*{8}{>{\centering\arraybackslash}p{19pt}}@{}}
\toprule
 & \multicolumn{2}{c}{\twoline{Program}{repair}} & \multicolumn{2}{c}{\twoline{Issue}{reproduction}} & \multicolumn{2}{c}{\twoline{Compiler}{autotuning}} & \multicolumn{2}{c}{\twoline{Static-analysis}{warnings}} \\
\cmidrule(lr){2-3}\cmidrule(lr){4-5}\cmidrule(lr){6-7}\cmidrule(l){8-9}
Criterion & \scriptsize Ours & \scriptsize Codex & \scriptsize Ours & \scriptsize Codex & \scriptsize Ours & \scriptsize Codex & \scriptsize Ours & \scriptsize Codex \\
\midrule
Delivered  & 50 & 50 & 50 & 50 & 50 & 50 & 48 & 50 \\
Real       & 50 & 50 & 49 & 45 & 50 & 50 & 48 & 49 \\
Admissible & 50 & 29 & 43 & 45 & 49 & 46 & 46 & 41 \\
Unique     & 50 & 28 & 43 & 43 & 49 & 14 & 46 & 36 \\
Complete   & 50 & 27 & 43 & 34 & 49 & 10 & 46 & 20 \\
Builds     & 50 & 12 & 43 & 7 & 49 & 6 & 46 & 5 \\
Functional & 49 & 5 & 43 & 3 & 49 & 6 & 45 & 0 \\
Expected   & 47 & 3 & 39 & 3 & 49 & 4 & 45 & 0 \\
\bottomrule
\end{tabular*}
\end{table}

Codex delivers 200 instances in 48 hours averaging 14 minutes and \$0.07 per instance. However, only 10 (5\%) of them meet all criteria. In comparison, of the 198 instances that BenchMeUp delivers, 180 (91\%) meet all criteria. Codex chooses sources (e.g., project names) from its knowledge without a search procedure; only 121 of its instances are real, admissible, and unique. Only 30 of its 91 complete instances build.
In contrast, the criteria that BenchMeUp checks during construction hold for every admissible instance: i) all are unique, ii) complete, and iii) build. These are the properties that deduplication, the construction plan, and replay in a fresh container are designed to ensure. They are also the criteria at which Codex fails most often.

The remaining BenchMeUp failures occur almost only at the specification's filters and at the expected outcome. Both depend on how the agents interpret the specification. Codex fails the filters more often than BenchMeUp. Most of BenchMeUp's failures occur in issue reproduction, where most instances that fail the filters come from projects with fewer than the 20 merged pull requests that the specification requires. In the issue-reproduction instances without the expected outcome, the test of the issue is not fail-to-pass.

\section{Discussion and Threats to Validity}
\label{sec:discussion}

\paragraph{Execution in a fresh container.}
In BenchMeUp, the framework checks stage completion by replaying the agent's scripts in a fresh container. This check is necessary: in 108 of 198 constructed instances, a stage that worked in the agent's container failed in the fresh one, most often because a build relied on something the agent had installed or left behind (Section~\ref{sec:rq1}).
Therefore, agent-based benchmark-construction approaches should re-execute every instance in a clean environment rather than trust the agent's report.
The instances' execution environments can decay as base images, packages, and repositories change; however, users can preserve environments through container snapshots (e.g., DockerHub).

\paragraph{Manually created specifications.}
The authors' knowledge of the approach may have made their four specifications easier for it to process than specifications by other researchers.

\paragraph{Choice of construction model.}
Our evaluation uses Claude Opus~5, which is stronger than deepseek-v4-flash, to generate search rules and construction plans and conduct searches, leaving repetitive plan execution to the weaker model. To investigate whether construction also benefits from a stronger model, we run Opus~5 on the same candidates for three tasks (excluding fixing static-analysis warnings for budget reasons). Opus~5 also delivers 50 instances per task at about \$10 per instance, but takes only about 9 minutes per instance, compared with 25 for deepseek-v4-flash on the same tasks. Opus~5 constructs instances from 40 candidates that deepseek-v4-flash drops, reducing wasted candidates and time. Hence, a stronger model can help with challenging tasks where a weaker one delivers fewer instances.

\paragraph{Limits of agent-generated checks.}
Replay guarantees that an instance reproduces the outcomes required by its checks. However, replay does not establish the correctness of these agent-written checks. Moreover, the checks reflect only the specification and the plan, i.e., they measure reproducibility and consistency with the plan, not faithful representation of the task.
Before using a benchmark, we recommend inspecting sample instances, as in RQ3.

\paragraph{Generalizability.}
We evaluate the approach on four tasks across three languages, using only GitHub data, with one specification and 50 requested instances per task. The approach is task-independent by design, but these results do not establish its effectiveness on tasks requiring other sources, such as non-GitHub issue trackers or app stores.

\section{Related Work}
\label{sec:related}

Table~\ref{tab:related} compares BenchMeUp with related approaches.

There are hundreds of software-engineering benchmarks~\cite{hu2025assessing}, most of which are curated for one task and one ecosystem.
HumanEval~\cite{chen2021humaneval} and MBPP~\cite{austin2021mbpp} evaluate code generation on small hand-written or crowd-sourced problems. Defects4J~\cite{just2014defects4j}, Bears~\cite{madeiral2019bears}, BugsInPy~\cite{widyasari2020bugsinpy}, Bugs.jar~\cite{saha2018bugsjar}, and GitBug-Java~\cite{silva2024gitbugjava} collect real bugs for program repair, and Vul4J~\cite{bui2022vul4j} takes a proof-of-vulnerability test as the correctness condition. SWE-bench~\cite{jimenez2024swebench}, SWT-bench~\cite{mundler2024swtbench}, and Multi-SWE-bench~\cite{zan2025multiswebench} cover issue resolution and reproduction.
LiveCodeBench~\cite{jain2025livecodebench} counters contamination by collecting problems continuously.
Instead of suggesting a single new benchmark, our approach automates the construction of benchmarks for arbitrary software-engineering tasks.

\begin{table}[t]
\centering
\caption{BenchMeUp and related approaches. ``Any''= no explicit restriction in the design. Issue res.= issue resolution; Code gen.= code generation; Env.\ setup= environment setup; PRs= pull requests; J= Java, Py= Python; Executed= executable and verified by execution; Interface= common eval interface.}
\label{tab:related}
\footnotesize
\setlength{\tabcolsep}{2.5pt}
\begin{tabular}{@{}lllllcc@{}}
\toprule
Approach & Task & Lang. & Input & Instances & \rotatebox{90}{Executed} & \rotatebox{90}{Interface} \\
\midrule
BugSwarm~\cite{tomassi2019bugswarm}             & CI fail/fix & J, Py & CI builds & Real & \checkmark & \checkmark \\
GitBug-Actions~\cite{saavedra2024gitbugactions} & Bug fix & Any & Repositories & Real & \checkmark & \checkmark \\
SWE-smith~\cite{yang2025swesmith}               & Issue res. & Py & Repositories & Synthetic & \checkmark & \checkmark \\
SWE-rebench~\cite{badertdinov2025swerebench}    & Issue res. & Py & Issues, PRs & Real & \checkmark & \checkmark \\
SWE-Factory~\cite{guo2026swefactory} & Issue res. & Any & Issues, PRs & Real & \checkmark & \checkmark \\
SWE-Bench++~\cite{wang2025swebenchpp} & Issue res. & Any & Issues, PRs & Real & \checkmark & \checkmark \\
CodeBenchGen~\cite{xie2024codebenchgen}         & Code gen. & Py & Code snippets & Real & \checkmark & \checkmark \\
SetUpAgent~\cite{vergopoulos2025setupagent}     & Env.\ setup & Py & Repositories & Real & \checkmark & \checkmark \\
RepoLaunch~\cite{li2026repolaunch}              & Env.\ setup & Any & Repositories & Real & \checkmark & \checkmark \\
ExecutionAgent~\cite{bouzenia2025you}           & Env.\ setup & Any & Repositories & Real & \checkmark & \xmark \\
SEC-bench~\cite{lee2025secbench}               & Vuln.\ patch & C/C++ & CVE reports & Real & \checkmark & \checkmark \\
AutoBencher~\cite{li2025autobencher}            & LLM eval. & N/A & Desiderata & Generated & \xmark & \checkmark \\
BenchAgents~\cite{butt2024benchagents}          & LLM eval. & N/A & NL description & Generated & \xmark & \checkmark \\
\midrule
BenchMeUp                                       & Any & Any & NL spec. & Real & \checkmark & \checkmark \\
\bottomrule
\end{tabular}
\end{table}

Tools for collecting instances mine repositories and rerun their builds. PyDriller~\cite{spadini2018pydriller} extracts commits, diffs, and developers from Git repositories, which task-specific scripts then filter. BugSwarm~\cite{tomassi2019bugswarm} collects reproducible failing and passing builds from continuous-integration histories, and GitBug-Actions~\cite{saavedra2024gitbugactions} reruns GitHub Actions workflows locally to find bug fixes and exports a Docker image for each. These tools are tied to a source of instances, such as commit histories or CI builds, and to one type of instance, whereas our search agents take the source and the selection criteria from the specification.

Techniques for automatically setting up and testing repositories~\cite{bouzenia2025you,vergopoulos2025setupagent,milliken2025installamatic,hu2025repo2run} automatically install dependencies, build a project, and run its tests. EnvBench~\cite{eliseeva2025envbench} shows that this task remains hard: The best approach it evaluates configures only 6.69\% of the Python and 29.47\% of the JVM repositories.
Task-construction techniques~\cite{yang2025swesmith,badertdinov2025swerebench,guo2026swefactory,wang2025swebenchpp,lee2025secbench} build on such setups to generate or collect issue-resolution and vulnerability tasks with executable environments at scale.
SWE-Factory~\cite{guo2026swefactory} and SEC-bench~\cite{lee2025secbench} are most similar to our work as they also use agents to construct executable software benchmarks, focusing on issue-resolution and vulnerability tasks, respectively.
Unlike in BenchMeUp, they rely on manually designed, task-specific definitions of an instance, the roles of the agents, and a correctness oracle.
CodeBenchGen~\cite{xie2024codebenchgen} and AutoCodeBench~\cite{chou2025autocodebench} use LLMs to turn code into execution-based code-generation benchmarks. Our approach differs by collecting and preparing real-world software artifacts.
SWE-Gym~\cite{pan2025swegym}, R2E~\cite{jain2024r2e}, R2E-Gym~\cite{jain2025r2egym}, and SWE-Synth~\cite{pham2025sweSynth} construct environments with executable tasks for issue resolution or for training agents. Our approach covers more software-engineering tasks and derives construction procedures from researcher-written benchmark specifications.

Studies of SWE-bench find solution leakage and weak tests~\cite{aleithan2024swebenchplus} and show that a share of patches counted as correct fail the developers' tests~\cite{wang2026solved}. UTBoost~\cite{yu2025utboost} generates additional tests after the fact, and BenchGuard~\cite{tu2026benchguard} audits existing benchmarks. Our correctness conditions and replay act at construction time; as Section~\ref{sec:discussion} discusses, the adequacy of an oracle remains the researcher's judgment. 

\section{Conclusion}
\label{sec:conclusion}

We present BenchMeUp, an approach that builds executable software-engineering benchmarks from a natural-language specification.
Shared search rules and construction plans coordinate parallel agents, while replay checks their outputs in fresh containers before packaging instances behind a common interface.
On four tasks, 180 of 198 constructed instances meet all criteria, compared with 10 of 200 for a general-purpose agent.
We envision BenchMeUp as a useful tool for researchers who want to evaluate their techniques on new tasks, programming languages, or software versions without the overhead of manually constructing benchmarks.

\bibliographystyle{IEEEtran}
\bibliography{references,referencesMichael}

\end{document}